\documentclass[aps,pra,twocolumn,groupedaddress,showpacs,showkeys,superscriptaddress]{revtex4-1}

\usepackage{amsmath}
\usepackage{amssymb}
\usepackage{graphicx}
\usepackage{epstopdf}

\usepackage{subcaption}
\usepackage{mwe}

\usepackage{media9}
\usepackage{hyperref}

\usepackage{cancel}
\usepackage{tabularx}

\DeclareMathOperator{\sign}{sgn}

\begin{document}

\title{Inverse design couplers for the excitation of odd plasmonic pairs in thin semiconducting films}

\author{Marius Puplauskis}
\email[correspondence address: ]{Marius.Puplauskis@skoltech.ru}
\affiliation{Skolkovo Institute of Science and Technology, Moscow 121205, Russian Federation}

\author{Ivan A. Pshenichnyuk}
\affiliation{Skolkovo Institute of Science and Technology, Moscow 121205, Russian Federation}

\date{\today}

\begin{abstract}
A set of grating couplers that convert plane waves into odd pairs of plasmons supported by extremely thin layers of doped indium tin oxide is designed. The inverse design approach is implemented to improve the efficiency of ordinary periodical couplers significantly. The optimization code based on the gradient descent method and direct Maxwell solver is designed in Matlab. Three models with different optimization levels are compared and discussed. The results of calculations are prepared for subsequent experimental verification. Considered odd plasmonic pairs represent an example of high quality modes tunable using a field effect. They are potentially applicable for the implementation in various active electro-optical devices and lead the way to fully semiconductor based plasmonics.
\end{abstract}

\maketitle

\section{Introduction} \label{sec_intro}

The concept of quasiparticles based electronics and photonics promises significant improvements in future technologies. The amount of research works devoted to the implementation of hybrid light-matter particles steadily grows \cite{rivera-2020,pshenichnyuk-2019b}. Impressive results are achieved with exciton-polaritons \cite{kavokin-2010} and plasmon-polaritons \cite{stockman-2011}, as well as with their combinations \cite{zhang-2020}. Exotic properties of exciton-polaritons, including their ability to form condensates at room temperature\cite{plumhof-2014}, are promising for both fundamental and applied science \cite{deng-2010,liew-2010,flayac-2013}. Their nonlinear behavior typical for Bose condensates, including their ability to guide solitons\cite{smirnov-2014} and quantum vortices \cite{pshenichnyuk-2017,pshenichnyuk-2015}, opens new frontiers in opto-electronics. The process of integration of plasmon-polaritons into modern opto-electronics already enters the industrial phase. Numerous plasmonic improvements to the existing photonic devices are suggested \cite{stockman-2018,pshenichnyuk-2019,pshenichnyuk-2018c}.

One of the major factors that inhibit the development of applied plasmonics is related to ohmic losses. The presence of metals causes attenuation of optical signals. There are several possibilities to overcome the difficulties using plasmonic amplifiers \cite{nezhad-2004,noginov-2008} and hybrid circuit design approaches \cite{alam-2013,alam-2014}. Another perspective idea is to develop fully semiconductor based plasmonics. Good tunability via doping and in general lower density of charges in semiconductors (to compare with noble metals traditionally used in plasmonics) allow to keep losses under control. At the same time, charge density profiles in semiconductors can be significantly influenced using a field effect. It potentially allows to use them as a natural platform for active plasmonic devices like electro-optical transistors and modulators. On the other hand, plasmonic properties in semiconductors are less pronounced than in metals, and the practical implementation of the concept requires the usage of advanced plasmonic mechanisms.

One possible realization of active plasmonics based on indium tin oxide (ITO) is described in the recent work \cite{pshenichnyuk-2021}. It is demonstrated there that an accumulation layer formed at a boundary ITO/dielectric can support surface plasmon polariton (SPP) modes. Accumulation layers are in general rather thin (few nanometers) and charge distribution profiles are highly inhomogeneous there. It has an influence on the field distribution profile of corresponding plasmons. Another feature that makes such thin film surface plasmon (FSPP) modes special is that they actually represent a coupled plasmonic pair. Similar to metallic films, odd plasmonic pairs propagate significantly further that single SPP, especially when the thickness of the layer becomes small \cite{novotny}. This fact guarantees large quality factors of FSPP modes and resolves the problem of poor propagation length of ordinary SPP modes in ITO \cite{naik-2013}. The ability to create and annihilate the accumulation layer using the applied fields makes FSPP modes at ITO/dielectric interface highly tunable and suitable for applications in active opto-electronic devices. It should be stressed here that the considered mechanism to control FSPP in ITO is qualitatively different from epsilon near zero (ENZ) effect often used for the realization of switching behavior in ITO \cite{pshenichnyuk-2021}. 

Along with the mechanism described above, a slightly modified principle suitable to control FSPP modes in ITO and available for immediate experimental verification was suggested in the work \cite{pshenichnyuk-2021}. Thin FSPP supporting layer of ITO with appropriate density can be originally prepared using doping. In contrast with the accumulation layer, it does not require the external voltage to be formed. Field effect in this case is used to distort the concentration profile and make the existence of the plasmonic mode impossible (referred as a 'destructive approach' in \cite{pshenichnyuk-2021}). One significant advantage of this approach is the possibility to control the thickness of the original FSPP carrying layer that has a direct and strong impact on the size of FSPP.

To verify the described effect experimentally it is necessary to excite an odd plasmonic pair in a thin layer of doped ITO. Then the external voltage can be used to manipulate the mode. The straightforward way to do that is to use a grating coupler, that coverts a plane wave into FSPP. A rather special character of the mode (a micron size double plasmon is supported by a few nanometers thick layer) leads to a small coupling efficiency of traditional gratings with constant period. For this reason the inverse design \cite{molesky-2018,piggott-2015} approach is implemented to maximize the coupling efficiency. This approach is demonstrated to be efficient for the significant improvement of grating couplers \cite{michaels-2018,su-2018}. To the best of our knowledge, it was not applied for the realization of coupler described here.

\section{Model}   \label{sec_modelintro}

\begin{figure}
	\centerline{\includegraphics[width=0.50\textwidth]{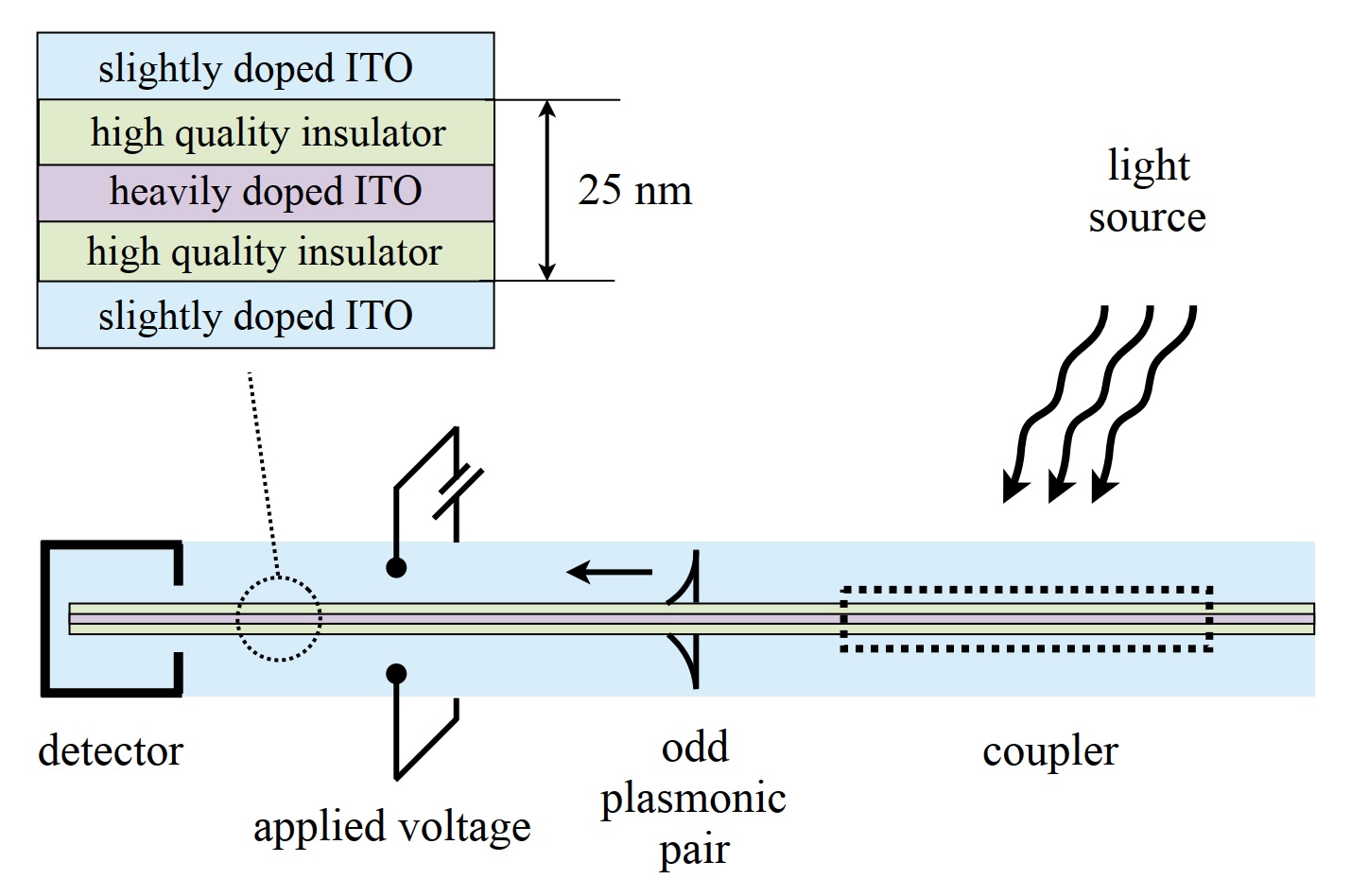}}
	\caption {Scheme of the proposed experimental setup for the excitation and detection of tunable plasmonic modes in thin semiconducting films. \label{experiment}}
\end{figure}

The proposed experimental setup for the excitation and detection of tunable odd plasmonic pairs in thin semiconducting films is presented in Fig.~\ref{experiment}.  The structure of FSPP carrying sandwich is discussed in details in our previous work \cite{pshenichnyuk-2021}. The sandwich includes a $5$ nm thick layer of heavily doped ITO. The concentration of charges $10^{21}$ cm$^{-3}$ is sufficient to allow SPP modes in such a material. ITO is isolated from both sides by $10$ nm thick high quality insulating layers of HfO$_2$, designed to keep the concentrated electron gas inside. The outer part of the sandwich is made of thick layers of ITO with relatively low density of charges ($10^{19}$ cm$^{-3}$). This sandwich operates as a double capacitor \cite{pshenichnyuk-2019}. The concentration of charges in the outer part is balanced to minimize the attenuation of the propagating plasmonic mode at the experimental length scale.

The dispersion relation of FSPP modes is presented and analized in \cite{pshenichnyuk-2021}
\begin{equation}
\beta^2 = \frac{\omega^2}{c^2}\varepsilon_{cl} + \frac{\eta^4}{2\eta^2 + 4\xi^2},
\label{fspp_disp}
\end{equation}
\begin{equation}
\eta \equiv \frac{\omega}{c}\sqrt{\varepsilon(\omega)-\varepsilon_{cl}}, \,\,\,
\xi \equiv \frac{1}{d}\frac{\varepsilon(\omega)}{\varepsilon_{cl}}.
\end{equation}
Here $\beta$ is the propagation constant of an odd plasmonic pair, $d$ - thickness of the active layer, $\varepsilon_{cl}$ - permittivity of the outer part (claddings), $\varepsilon(\omega)$ - plasma dominated permittivity of the active layer. The behavior and quality factors of FSPP strongly depend on $d$ and the wavelength of light $\lambda_0 = 2\pi{c}/\omega$. In the experimental design proposed here we use $d=5$ nm, $\lambda_0 = 957$ nm. Thickness $d$ should be small enough to allow effective tuning using an external voltage. It can not be too small, because of the experimental accuracy limitations. The wavelength of light is also well balanced here for the experimental verification of the effect. There are no FSPP modes available for smaller wavelengths and for larger wavelengths the transverse size of FSPP grows fast and becomes experimentally impractical. The expected quality factor of the considered FSPP mode is close to $12$ and losses are about  $0.26$ dB/$\mu$m. Plasmons with such characteristics can be easily detected. The expected voltage, necessary to manipulate the mode, should be close to $30$ V \cite{pshenichnyuk-2021}.

The only significant problem with the experimental verification of the effect is related to the low efficiency of excitation of such modes using grating couplers. According to our simulations, the coupling efficiency of an ordinary coupler with a constant period in this case is close to $-47$ dB, which is quite hard to detect and analyze experimentally. The reason for that is a tiny width of FSPP carrying layer. Another reason is related to ohmic losses inside the coupler. The improvement of $5$ dB in this case can make a notable difference. That is what we aim to achieve here using the inverse design approach.

\section{Theory and methods}   \label{sec_theory}

\begin{figure}
	\centerline{\includegraphics[width=0.50\textwidth]{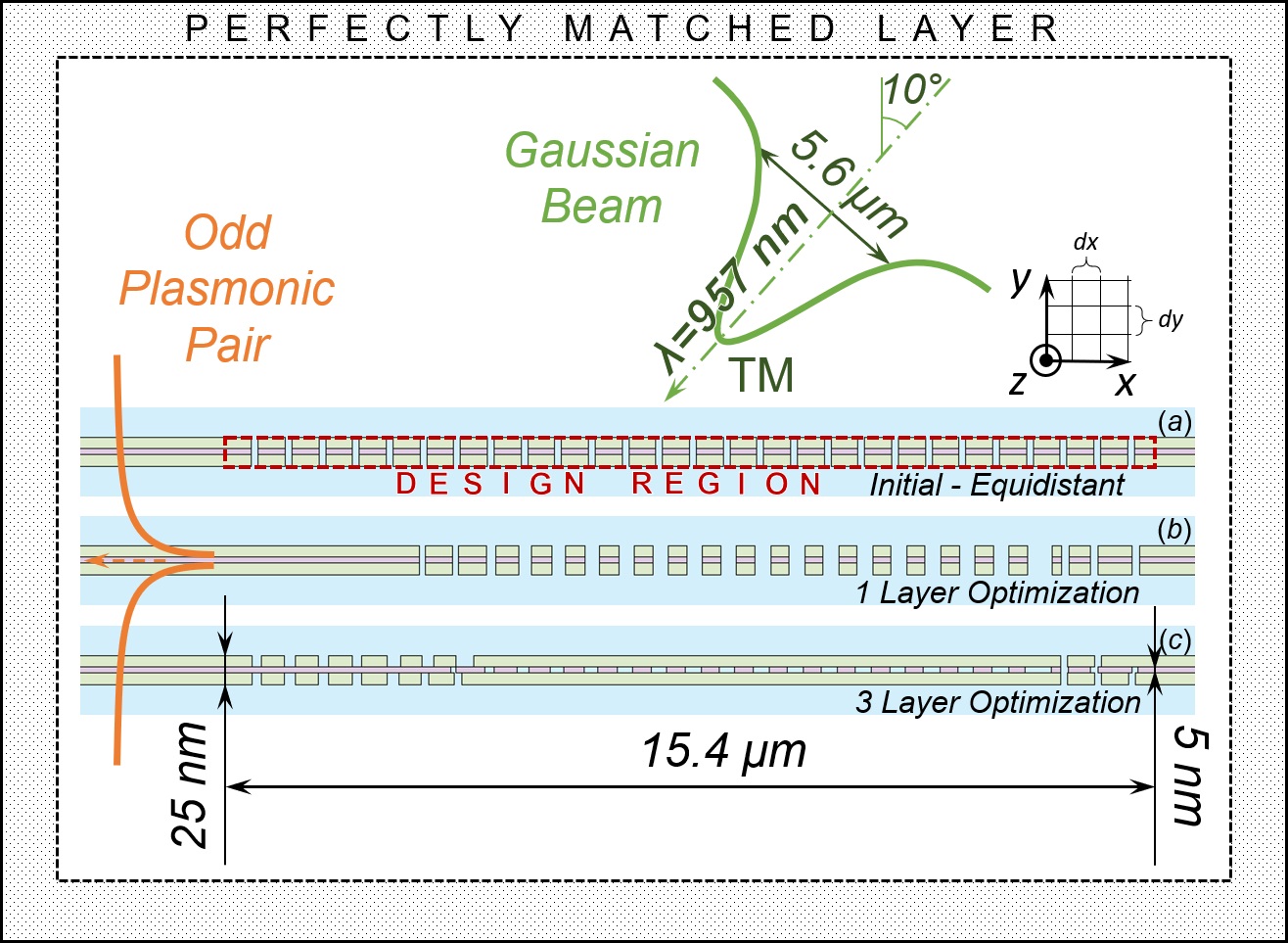}}
	\caption {Geometry of the original (a) and two inverse designed (b),(c) grating couplers. Parameters of the light source and axis orientation are shown. \label{Geometry}}
\end{figure}

Inverse design is an algorithmic technique for discovering optimal structures based on desired functional characteristics $d$ or, in other words, a desired data to be observed \cite{molesky-2018}. This data can be acquired using an observation function $F=F(\psi)$ or, as it is usually called, a forward mapping, applied to the field distribution $\psi=\psi(p)$ (electromagnetic field in our case). The field itself depends on parameters $p=p(r,c)$ (such as dielectric permittivity) defined over a spatial coordinate $r \in \mathbb{R}^n$ ($n$ is the number of spatial dimensions). Parameters $p$ vary within the predetermined constraints $c$. In our paper, for example, one of the constraints limits the domain where $p$ is varied to a subdomain $S \in \mathbb{R}^n$ that we call the optimization or design region. In the rest of the domain $\overline{S}$ parameters are fixed. Other constraints are introduced to guarantee that the optimized structure is fabricable and are described further in the text.

A forward problem is defined as an application of a forward mapping to calculate an observed data for some specific set of parameters $p$
\begin{equation}
	d_{obs} = F(\psi,p,r,c),
\end{equation}
while an inverse problem requires to determine the optimal set of parameters $p_{opt}$ that leads to the desired observation $d_{des}$ (in this paper we consider it as an electromagnetic power flow through a boundary)
\begin{equation}
	p_{opt}=F^{-1}(\psi,d_{des},r,c).
\end{equation}
In most practical cases $F^{-1}$ can not be found analytically. Moreover, numerical evaluation is usually computationally demanding.

One efficient way to handle the inverse problem numerically is the gradient descent method. It is designed  to minimize the difference between current and desired observation iteratively. In other words, a discrepancy between observed and desired data
\begin{equation}
	\mathcal{F}(p)= \| F(\psi,p,r,c)-d_{des} \|^m
\end{equation}
is minimized in order to find an optimal set of parameters $p_{opt}$. The discrepancy $\mathcal{F}$ is also referred to as an objective function or an objective functional. Here we present it in a standard form
with the norm order of $m=1$.

The gradient descent method implies that each new iteration of the algorithm leads to a smaller discrepancy. Solving an inverse problem can be mathematically formulated as a minimization of discrepancy over the parameters domain
\begin{equation}
	p_{opt}=\mathcal{F}^{-1} \Big[\min_{p}\big|F(p)-d_{des}\big|\Big].
\end{equation}
One critical part here is to choose a correct descent direction and step at each iteration. In an analytical case, if a functional form of the objective function gradient can be determined explicitly, the preferred direction of a descent is opposite to it
\begin{equation}
	n = - \delta \mathcal{F} / \delta p = -\delta_{p} \mathcal{F}.
\end{equation}
The descent step $h$ can either be some constant value or a function of speed, acceleration, and other parameters and limitations.

The gradient descent algorithm can be schematically described as follows:
\begin{enumerate}
	\item Domain $\overline{S}$ is set, values $h_0$, $p_0$ are initialized, $\mathcal{F}_0=\mathcal{F}(p_0)$ is calculated;
	\item Variation $\delta_{p} \mathcal{F}$ is calculated;
	\item New descent direction $n_j$ and step $h_j$ and thus parameter value $p_j$ are calculated $p_j=p_{j-1}+n_j\cdot h_j$;
	\item New objective function value is calculated and compared with the previous: $\mathcal{F}(p_j)  >/< \mathcal{F}_{j-1}$;
	\begin{itemize}
		\item If $\mathcal{F}_j>\mathcal{F}_{j-1}$ the algorithm might be stopped or returned to step 3 to recalculate descent direction and step;
		\item Else, if the desired objective value is not reached $\mathcal{F}_j>\mathcal{F}_{des}$ the algorithm continues back to step 2;
		\item Otherwise, the algorithm ends with $p_{opt}=p_j$. 
	\end{itemize}
\end{enumerate}
This algorithm is highly dependent on the calculation of the objective function variation $\delta_{p} \mathcal{F}$. Latter is obtained by solving a forward problem. Thus, an inverse problem solving procedure contains in itself a forward problem. It is the most computationally expensive part of the  algorithm. Roughly, its performance can be estimated as a ratio of the objective function decrease to the number of times the forward problem has to be solved.

Depending on how the variation $\delta_{p_i} \mathcal{F}$ is calculated several types of the algorithm can be distinguished. In this work we use the batch gradient descent method that implies the calculation of a variation for each parameter $p$. This approach provides a stable convergence and works well for the initial condition that we use (more details in the next section). Other types of gradient descent methods are more thoroughly described in \cite{ruder2016overview}.

In the numerical case the computational domain $\mathbb{R}^2 = x \otimes y$ is discrete and finite: $ x \otimes y \rightarrowtail {dx,dy} \in C_d \subset \mathbb{R}^2 $, where $C_d$ stands for the discrete domain and the design region is its subspace $D_r \subset C_d$ (see Fig.~\ref{Geometry}). A uniform grid is used in this work with $dx=const_x$, $dy=const_y$. Perfectly matched layers (PML) are used as a boundary conditions to imitate open boundaries. These layers strongly absorb outgoing waves from a computational region’s interior without reflecting them back. 

The discrete computational domain implies that instead of exact partial variations $\delta_p \mathcal{F}$, we calculate finite increments of the objective function $\Delta_p \mathcal{F}$ with an accuracy inversely proportional to the parameter increment $(\Delta p)^n, n \in N$. Naturally, parameters $p$ also form a discrete and finite set of size $m$. Discretization causes inaccuracies in the descent direction calculation. Additional challenge in our case comes from the fact that the size ratio between the largest element (grating and waveguide length) and the smallest one (thickness of FSPP carrying layer) in the computational domain is larger than three orders of magnitude. Therefore, a reasonable balance should be found between the performance of the algorithm and partial derivative's computation accuracy.

A partial sacrifice of the accuracy allows to use a coarser computational grid. For this reason, the finite differences of the parameters $\Delta p$ turn out to be equal or greater than the grid constant in the horizontal direction $dx$. To avoid the rise of errors in finite increment calculations, we introduce the following selection rules. All finite differences $\Delta \mathcal{F}_{p_i}, \bigcup p_i = p_{D_r}(r \in D_r )$ that are positive $\Delta \mathcal{F}_{p_i} > 0$ (since we are minimizing the functional) and exceed the minimum finite difference in absolute value $\Delta \mathcal{F}_{p_i} > \min_{i} |\Delta \mathcal{F}_{p_i}|$ are not considered when composing a vector of a descent direction. The reasons for such selection become clear after a full consideration of the descent step (see below).
It is essential to elaborate on notations used: $\Delta_p \mathcal{F} = \Delta \mathcal{F}_p / \Delta p $, where $\Delta_{p_i} \mathcal{F}$ – finite increment of the objective function, $\Delta \mathcal{F}_{p_i}$ – finite difference of objective function, $\Delta p_i$ – finite difference of parameter. The remaining finite differences of the objective function $\Delta_{p_i} \mathcal{F}$ are divided by the corresponding finite differences of the parameters $\Delta p_i$ that led to the rise of the function. The resulting vector is normalized using the maximal finite increment from the remaining subset
\begin{equation}
\label{eqn:n calc}
\begin{aligned}
	n &= - \bigg\{ \cfrac{ \Delta \mathcal{F}_{p_k} }{ \Delta p_k } \bigg\}_{k=\overline{1,m-l}} \bigg/ 
	\max_k \bigg|\cfrac{ \Delta \mathcal{F}_{p_k} }{ \Delta p_k }\bigg|, 
	\\ &\quad \big(\Delta \mathcal{F}_{p_k} < 0\big) \cap \big( \Delta \mathcal{F}_{p_k} < \min_i |\Delta \mathcal{F}_{p_i}| \big) \doteq K, i=\overline{1,m}
\end{aligned}
\end{equation}
where $l$ – is the number of excluded finite differences $\Delta \mathcal{F}_{p_i}$.

Before the discussion of constraints $c$ and descent step $h$ we would like to elaborate on how we map the dielectric permittivity distribution inside the design region to a finite set of parameters $p$. The latter is a matrix of size $a \times 2b = m$, where $a$ – is the number of layers to be optimized and $b$ – is the number of strokes in each layer. Strokes are assumed to be filled by the surrounding medium (shown by the light blue color in Fig.~\ref{Geometry}). Layers thicknesses remain fixed during the optimization and each stroke in a corresponding layer is defined by two $x$-coordinates (beginning and ending).

Having in mind that the smallest parameter difference cannot be smaller than a grid constant in a horizontal direction $\min_i \Delta p_i \geq dx$ we choose $ |\Delta p_i| = dx $ when calculating the objective function finite increments $\Delta_{p_i} \mathcal{F}$. This ensures the most accurate descent direction calculation for the selected grid meshing. The parameter differences $\Delta {p_i}$ are chosen randomly at each iteration in the direction along the horizontal $x$ axis. Therefore, calculation of the descent direction vector $n$ in (\ref{eqn:n calc}) can be rewritten as:
\begin{equation}
	n=-\cfrac{\Delta \mathcal{F}_p}{\max |\Delta \mathcal{F}_p|} \cdot \sign (\Delta p), \Delta \mathcal{F}_p \in K.
\end{equation}
With the previous reasoning in mind, we conclude that the most optimal (in the sense of overlooking a possible local minimum) and, at the same time, straightforward approach in choosing a descent step $h$ would be a sweep along a gradient direction from some minimum to a maximum value of a step. Thus, a new objective function value at iteration $j$ is obtained as
\begin{equation}
	\mathcal{F}_j = \min_h \mathcal{F}(p_{j-1} + n_j \cdot h).
\end{equation}  
With a maximum value of $n_k$ being normalized to unity, minimum sweep value is equal to a horizontal grid constant $const_x$. The sweep step value must be a multiple of latter and we choose it to be equal to it. The maximum step can be an order of magnitude greater than the minimum value since the computational "bottleneck" is the evaluation of a finite differences $\Delta \mathcal{F}_p$.

It may turn out that during the sweep over a descent step, there is no such a value of the objective function that is smaller than its value from the previous iteration $\mathcal{F}_j \cancel{<} \mathcal{F}_{j-1}$. There may be several reasons for this. The first one is that the strokes positions are not independent. Even though all individual changes $\Delta p_k$ along the descent direction decrease the value of the objective function $\Delta \mathcal{F}_{p_k} < 0$, the joint change $\Delta p$ may lead, on the contrary, to an overall increase $\Delta \mathcal{F}_p > 0$. The second possible reason is related to the fact that we considered both negative and positive finite differences of the objective function $\Delta \mathcal{F}_{p_k}$ when composing the vector of descent direction $n$. Positive differences impose that an opposite direction of a descent must be chosen for a corresponding parameter $p_k$, but this does not guarantee a decrease of an objective function $\Delta \mathcal{F}_{p_k}$. Such behavior is typical near local minima. In order not to recalculate finite variations in cases of positive differences $\Delta \mathcal{F}_{p_k} > 0$, we utilize the selection process described above. One may think of it as a simplified weighing. By prioritizing precisely known functional decrease directions when composing the descent direction vector $n$ the weighing reduces the probability of "overshooting" a local minima. 

If the described selection still turns out to be insufficient to achieve a new smaller value of the objective function $\mathcal{F}_j \cancel{<} \mathcal{F}_{j-1}$, then the algorithm proceeds to eliminate the mutual influence of parameters. First, all positive finite differences are excluded from the decomposition of the descent direction vector $n$. Second, negative differences are excluded in a loop, starting from the smallest one (in an absolute value). With each new exclusion, the direct problem is solved with a descent step equal to the minimum value of the step $h$ sweep.

As it was already mentioned, constraints $c$ also regulate the values that the parameters $p$ and their differences $\Delta p$ can take inside the design region $D_r$. Namely, parameter values must be multiples of the grid constant in the horizontal direction $dx$; parameters should not take values exceeding the design region boundaries; the distance between an adjacent positions of the strokes edges should not be less than a given value. This value is determined by the technical feasibility of a manufacturing geometry with a given resolution. For the optimizations presented in this article, it is taken to be equal to $100$ nm. If the absolute value of the finite difference of some parameter $|\Delta p_i|$ violates the last condition, then it is set to the value for which two competing strokes merge. If this change leads to an overall decrease of the objective function $\mathcal{F}$, then these two strokes are considered as one in the following iterations. The reverse process of "emergence" of a "tooth" between stroke is not provided by the algorithm. According to our results (summarized in the next section) final geometries do not allow even a single "emergence" event without a violation of constraints $c$.

Electromagnetic fields $\psi$ used to calculate the observed data $d$ are governed by the set of Maxwell’s equations. The frequency domain version of the equations is used in this work since we are interested in a steady state solution. 2D computational domain is considered to keep the required computational resources under control. The resulting model represents the situation where the grating coupler and a waveguide are sufficiently wide. The sketch of a 2D slice of the forward problem computational domain is presented in Fig.~\ref{Geometry}.

An efficient grating coupler that we design in this paper is supposed to convert as much of the impending electromagnetic power as possible into the power of an odd plasmonic pair. We define the desired data $d_{des}$ as an overall power outflux (Poynting vector) from the computational domain $C_d$ in the form of odd plasmonic pair traveling along with a metal-like and dielectric-like interface of semiconductors (see Fig.~\ref{Geometry}). 

The described algorithm is realized in MATLAB. The external library for Maxwell solver is used \cite{maxwellfdfd-webpage}. The results are verified in COMSOL Multiphysics. All computational domain parameters are presented in the table below \ref{tab_param}. 

\begin{table}[h!]
\centering
\noindent\begin{tabular}{ |p{5.2cm}||p{2.3cm}|p{0.75cm}| }
	\hline	
	\textbf{Parameter} & \textbf{Value} & \textbf{Unit} \\	
	\hline \hline	
	Refractive index of ITO           		 	& $1.97 – i\,7\cdot10^{-4}$	& $1$ 	\\
	Refractive index of doped ITO		 	 	& $0.33 – i\,1.25$ 			& $1$ 	\\
	Refractive index of HfO\textsubscript{2} 	& $1.94$ 					& $1$ 	\\
	Computational domain width 			 	& 21000 					& nm 	\\
	Computational domain height 			& 500 						& nm 	\\
	PML thickness 							& 200 						& nm 	\\
	Waveguide (buffer) length 				& 5570 						& nm 	\\
	Grating (design region) length 			& 15430 					& nm 	\\
	Insulating layer thickness 				& 10 						& nm 	\\
	Doped ITO layer thickness 				& 5 						& nm 	\\
	Incident beam diameter, $D$ 			& 7354 						& nm 	\\
	Incident beam angle, $\alpha$ 			& 10 						& deg 	\\
	Wavelength of light, $\lambda$  		& 957 						& nm 	\\
	Period of the initial grating, $T$ 		& 585 						& nm 	\\
	Fill factor of the initial grating 		& 0.2 						& 1		\\
	\hline
\end{tabular}
\caption{Parameters used in the simulation}
\label{tab_param}
\end{table}

\section{Results and discussion}  \label{sec_results}

It is assumed that the impeding electromagnetic radiation comes from an optical fiber. Gaussian beam is a good theoretical approximation of such a light source. We model it as a set of point sources (dipoles) of an electric field aligned along a straight horizontal line located at the top boundary of the computational domain. In order to manipulate the angle of incidence phase shifts are applied according to the formula:
\begin{equation}
	E(x) = \sum_i E_0 \cdot \exp\bigg(-\cfrac{(x_i-x_0)^2}{2\sigma^2} + i\cfrac{2\pi}{\lambda}x_i\sin\alpha \bigg),
\end{equation}
where $x_0$ – position of the emitter in a horizontal direction, $\alpha$ – angle of incidence of the beam, $\lambda$ – wavelength of the light source. We define the beam diameter as a distance between points where the intensity $I=|E(r)|^2$ falls to $1/e^2  = 0.135$ times the maximal value. For a parameter $\sigma = 2.8$ $\mu$m we acquire the beam diameter of $7354$ nm.

\begin{figure}
	\centerline{\includegraphics[width=0.50\textwidth]{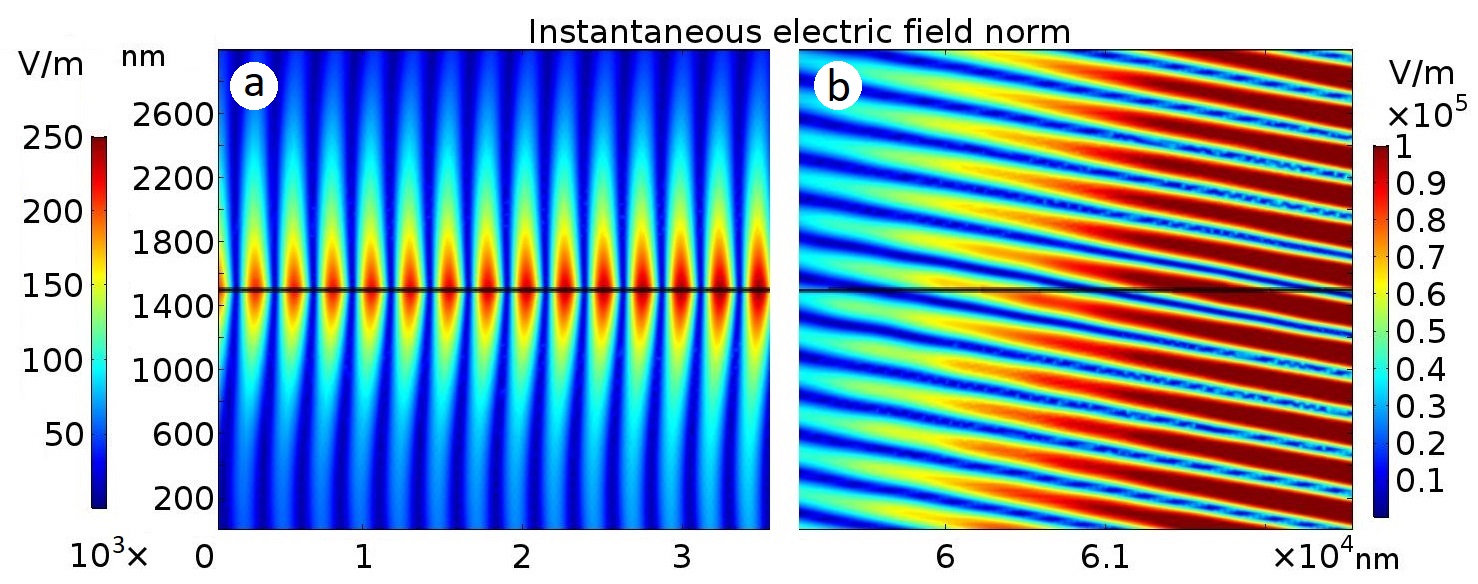}}
	\caption { Two modes to be coupled via grating (a) Odd plasmonic pair and (b) Gaussian beam from an optical fiber. A normalized electric field is color coded. FSPP supporting active layer is shown using horizontal black line.\label{Fields}}
\end{figure}

Impending on the grating, the beam excites an odd plasmonic pair, as shown in Fig.~\ref{Fields}a. This mode is characterized by a real part of the effective index $\Re (n_{eff})=1.9773$. Based on that number, the initial period of the grating coupler is evaluated as $T= 585$ nm. It is chosen so to compensate the lack of (horizontal) $x$-component of the incident wave vector as follows:
\begin{equation}
	\begin{aligned}
		k_x + k_{gr} = k_{wg} & \rightarrow \\ \cfrac{2\pi}{\lambda} n^\Re_{ITO_{19}}&\sin\alpha + \cfrac{2\pi}{T} = \cfrac{2\pi}{\lambda} n^\Re_{eff} \rightarrow \\ & T = \lambda/( n^\Re_{eff}  - n^\Re_{ITO_{19}} \sin\alpha)
	\end{aligned}	
\end{equation}
We also use the fill factor $0.2$ for the initial condition. The initial geometry is shown in Fig.~\ref{Geometry}a.

\begin{figure}
	\centerline{\includegraphics[width=0.50\textwidth]{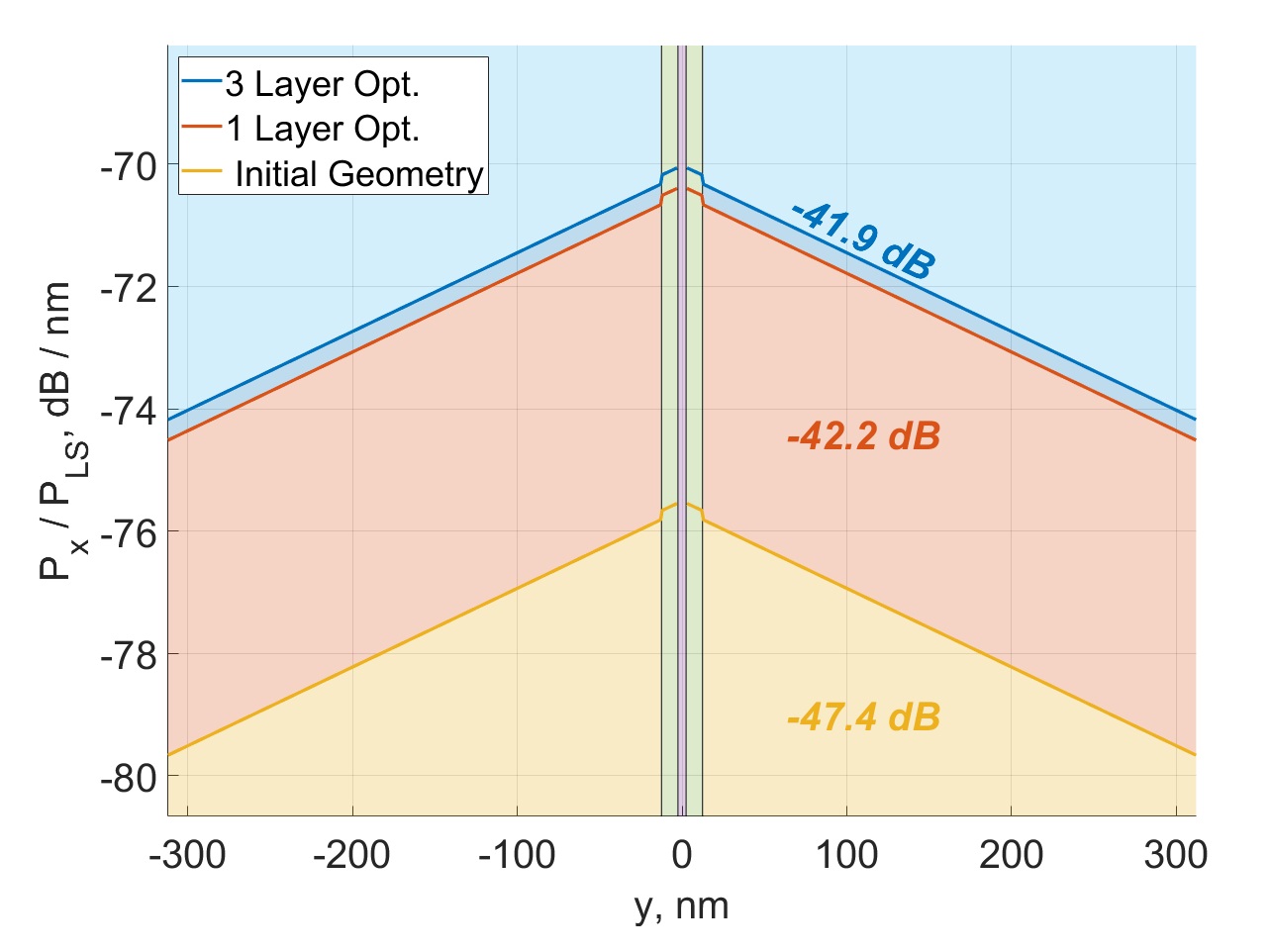}}
	\caption {Horizontal $x$-component of the energy flux density $P_x$ (along the interface) normalized by the total power of the light source $P_{LS}$. Integral values are shown with corresponding colors. \label{Power Flow}}
\end{figure}

We apply our algorithm to perform two types of the optimization. In the first case there is just one optimization layer in the design region (Fig.~\ref{Geometry}b). It includes both highly doped ITO and two insulating layers of HfO$_2$. Thus each independent period of the inverse designed grating is composed from three physical layers. In the second case (Fig.~\ref{Geometry}c) we split the optimization region into three layers and allow the algorithm to optimize them independently.
To evaluate the coupling efficiency we monitor the energy flux density at the output of the model and compare it with the power of the source, as it is shown in Fig.~\ref{Power Flow}. Energy density of a plasmonic pair goes down exponentially (linearly in the logarithmic scale that we use) with the distance from the interface. FSPP modes inside the interface demonstrate additional fine structure of the field \cite{pshenichnyuk-2021}. Integral values representing the total coupling efficiency are shown using corresponding colors.
We managed to increased the plasmon excitation efficiency approximately $4$ times, from $1.8 \cdot 10^{-3}\%$ for the initial equidistant grating to $6.0 \cdot 10^{-3}\%$ using the single-layer optimization and up to $6.4 \cdot 10^{-3}\%$  for the three-layer optimization.

\begin{figure}
	\centerline{\includegraphics[width=0.50\textwidth]{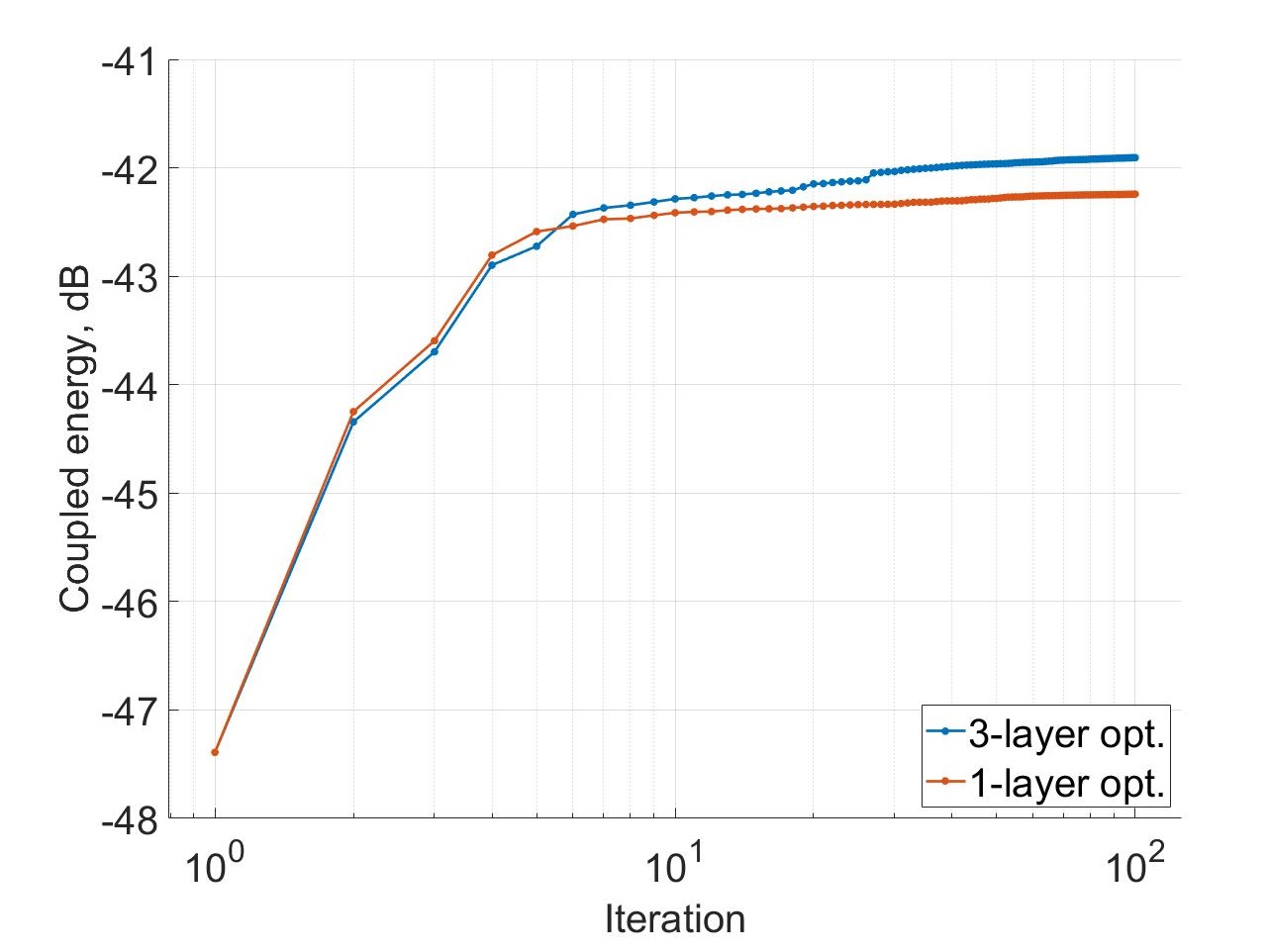}}
	\caption {The optimization process for one-layer (orange line) and three-layer (blue) models. One hanged iterations are shown. \label{Optimization}}
\end{figure}

The optimization process for both single-layer and three-layer models is visualized in Fig.~\ref{Optimization}. The algorithm well converges after approximately $100$ iterations. Computations are performed using Intel Core i5-7400 CPU at 3.00 GHz and 8 GB of RAM. An average time spent on one iteration is close to $16$ minutes. Thus, the problem is solvable using an ordinary PC or laptop. The number of parameters varied during the optimization is equal to $54$ for a single layer. Average time per iteration and number of parameters have to be multiplied by the number of layers in a case of multi layer optimization.

Qualitative changes of the geometry, introduced by the optimization algorithm, are shown in Fig.~\ref{Geometry}.  Strokes in the active layer tend to merge at the left hand side of the grating. Then their length increases and remains almost constant for the most of the grating length. They become more narrow again at the right hand side of the grating. There is also one twice-as-long stroke in the transition region. This pattern is typical for both single-layer and three-layer optimizations.
When we allow the strokes in the insulating layers to vary independently from the strokes in the active layer (three-layer optimization model), the tendency for central layer to increase the strokes length with its horizontal coordinate increase persists. However, in the insulating layers, a strokes merging is observed in the middle of the grating instead of its left hand side. Insulating layer's strokes at the left hand side are not much affected by the optimization. The insulating layers are qualitatively symmetrical, but there are some differences. Thereby, a larger increase in the strokes period and length with horizontal coordinate is observed in the upper layer. In the lower layer, strokes lengths demonstrate a concave down parabolic dependence on the coordinate. However, the merging of strokes in both insulators is symmetrical.

Analyzing the variations in the geometry of gratings (not presented) throughout the optimization process (Fig.~\ref{Optimization}) we conclude that the main contribution to the efficiency is due to changes in the active layer of highly doped ITO. Only after approximately fifteen iterations (when single-layer optimization has already converged) independent changes in the insulating layers start to take place. These changes provide an additional moderate increase in the efficiency when compared with single-layer optimization (compare orange and blue lines in Fig.~\ref{Optimization}). This result is expected since the refractive index of the insulator HfO$_2$ ($1.94$) considered here is quite close to the refractive index of the slightly doped ITO in the outer layer ($1.97$). From the optical point of view the difference between the strokes material and grating material is very weak and the resulting insulating layers gratings are quite 'soft'. Thus, the optimization of the insulators may provide only a slight boost to the efficiency. 
Moreover, the three layers optimization significantly increase the complexity of subsequent manufacturing of the coupler. The price to be payed for additional $0.3$ dB in the experiment may be too high and we suggest to use the single layer model for the subsequent lab tests.

\section*{Conclusion}  \label{sec_conclusion}

In this work we suggest a model of the efficient grating coupler for the conversion of light emitted from an optical fiber into a tunable odd plasmonic pair supported by a thin semiconducting film. An inverse design approach is implemented to maximize the efficiency of the coupler. The optimization algorithm based on the gradient descent method is realized in Matlab. Several qualitative improvements of the standard algorithm are implemented, including the selection rules and the handler of strokes merging events. The resulting algorithm is effective enough to be implemented using ordinary computers and laptops.

Two types of couplers are designed using one-level and three-level optimization. The latter one provides larger efficiency but it is much harder to realize experimentally. One-layer inverse design coupler in general provides a significant (more than $5$ dB) improvement over the ordinary constant period coupler. It can be easily manufactured using standard lithographic techniques. The designed coupler can be implemented experimentally for the subsequent investigation of tunable FSPP modes in semiconductors.

\bibliography{paper}

\end{document}